\documentclass[epj]{svjour}
\usepackage{graphics}
\usepackage{amssymb}
\usepackage{amsmath}

\newcommand \phe{$\phi\rightarrow e^{+}e^{-}$ }
\newcommand \phk{$\phi\rightarrow K^{+}K^{-}$ }
\newcommand \ee{$e^+e^-$ }
\newcommand \kk{$K^+K^-$ }
\newcommand \mee{$m_{e^{+}e^{-}}$ }
\newcommand \sqn{$\sqrt{s_{_{NN}}}$ }
\newcommand \gvce{$GeV/c^2$ }
\newcommand \mvce{$MeV/c^2$ }
\newcommand \gvcp{$GeV/c$ }
\newcommand \dau{$~d+Au$ }
\newcommand \au{$Au+Au$ }

\begin{document}
\title{Low-mass dielectrons from the PHENIX experiment at RHIC}
\author{Alexander Kozlov\inst{1} for the PHENIX collaboration
}
%
%
\institute{Weizmann Institute of Science, Rehovot 76100, Israel}
\date{Received: 15.08.2006 / Revised version: 15.08.2006}
%
\abstract{ The production of the low-mass dielectrons is considered to be a powerful tool to study the properties of the
  hot and dense matter created in the ultra-relativistic heavy-ion collisions. We present the preliminary results on the
  first measurements of the low-mass dielectron continuum in \au collisions and the $\phi$ meson production measured in
  \au and \dau collisions at \sqn = 200 $GeV$ performed by the PHENIX experiment.
  \PACS{{25.75.-q}{Relativistic heavy-ion collisions} \and {12.38.Mh}{Quark-gluon plasma}} 
 } 
\maketitle
\section{Introduction}
\label{intro}
Among the many diagnostic tools for the hot and dense matter produced in high energy heavy-ion collisions the production
of low-mass lepton pairs\footnote{Low-mass dilepton pairs include the light vector mesons and the
dielectron continuum with mass $m_{e^+e^-}\lesssim$ 1 \gvce.} plays an important role. Dileptons are emitted during the
entire lifetime of the collision and interacting only electromagnetically, escape the interaction region almost
freely carrying information directly to the detector. This makes dileptons an excellent tool to study the thermal
radiation emitted by the dense medium and possible in-medium modifications of the light vector mesons properties (mass
and/or width). This modifications are considered important signals of the restoration of chiral symmetry.

The production of low-mass electron pairs was extensively explored by the CERES collaboration at the SPS~\cite{CERES}.
CERES discovered an excess of the dielectron yield in the mass range 0.2-0.6 \gvce as compared to the one
expected from the hadronic sources. A similar excess was observed for different colliding energies and species by
DLS at the BEVALAC~\cite{DLS}, E325 at KEK~\cite{E325} and recently by the second generation experiments NA60 at the
SPS~\cite{NA60} and HADES at GSI~\cite{HADES}. For a summary of the most recent experimental results
see~\cite{Tserruya:2006ht}. 

The absolute yield and shape of the low-mass dielectron spectra measured by CERES are described satisfactorily by
various theoretical models involving thermal radiation from the $\pi\bar{\pi}$ annihilation and in-medium modifications
of the $\rho$, $\omega$ and $\phi$ vector mesons spectral functions~\cite{THRev}.

The $\phi$ meson is considered to be a sensitive probe of the possible restoration of chiral symmetry in relativistic
heavy-ion collisions which could manifest itself in the modification of its spectral properties (peak position and/or
width) and in the changes of  the relative yield as measured through the \ee \textit{vs.} the \kk decay
channels~\cite{shur91,pal02}. Since $m_{\phi}$ is slightly larger than $2 m_{K}$, even small changes in the spectral
properties of the $\phi$ or $K$ can induce significant changes in the branching ratio of the \phk decay.  

The lifetime of the $\phi$ meson is long compared to the lifetime of the fireball at RHIC energies ($\tau_{\phi} \approx $
46 $fm/c$ \textit{vs.} $\tau_{fireball} \approx $ 10 $fm/c$ ~\cite{Harris:1996zx}). In this case only a small fraction
of the $\phi$ mesons will decay inside the fireball, which could lead to a distortion of the line-shape if the $\phi$
mesons spectral function is affected by the medium. For example, a low-mass tail should develop if the $\phi$-meson
mass decreases in the medium.

The PHENIX detector has the potential to measure accurately \ee pairs in the low-mass region and the light vector meson
properties. The observation of a $\phi$-meson shape distortion is difficult but could be possible with the excellent mass
resolution, of about 1\%, of the PHENIX detector in the \ee decay channel provided that a good signal to background, $S/B$,
ratio can be achieved. In addition to that, the PHENIX detector has the unique capability of to measure simultaneously the
$\phi$-meson production through the \ee and \kk decay channels.  

During RHIC run 4 in 2004, PHENIX has collected about 241 $\mu$b$^{-1}$ integrated luminosity which allowed us
to perform a first measurement of the low-mass dielectron continuum and the $\phi$-meson production via \ee and
\kk decay channels in \au collisions at \sqn = 200 $GeV$. 

In this paper the low-mass dielectron mass spectra measured in \au collisions at \sqn = 200 $GeV$~\cite{HQtoia} and the
results of the $\phi$-meson production in \dau~\cite{seto} and \au~\cite{QMak} collisions at \sqn = 200 $GeV$ as
determined by the \ee and \kk channels are presented.

The measurement of low-mass \ee pairs with the present PHENIX configuration is challenging due to huge
combinatorial background originating from unrecognized conversions and $\pi^0$ Dalitz decays. Indeed, the $S/B$
ratio measured in \au collisions at \sqn = 200 $GeV$ is of the order of $\sim$~1/150 and $\sim$ 1/60 at \mee $\approx$
400 \mvce \cite{HQtoia} and $\phi$ mass~\cite{QMak}, respectively.

A PHENIX upgrade with a Hadron Blind Detector foreseen in 2006 will significantly reduce the combinatorial background,
improving the capability of PHENIX to measure low-mass dielectron pairs in heavy-ion collisions. The details of the HBD
project are presented below. 

\section{The PHENIX experiment}
\label{sec:1}

\subsection{Experimental apparatus}
\label{sec:1a}
PHENIX is one of four experiments at the Relativistic Heavy Ion Collider (RHIC) at Brookhaven National Laboratory.
Among them, PHENIX is the only one which has the capability to measure low-mass electron pairs. 
The PHENIX spectrometer~\cite{Adc01} consists of two central arms which cover a pseudorapidity range
$\vert\eta\vert <$ 0.35 and 2 $\times$ 90$^{\circ}$ in azimuthal angle. The momentum and charge of the particles
are determined using a Drift Chamber (DC) and a Pad Chamber (PC1). Electrons are identified by a Ring Imaging
Cherenkov (RICH) detector and by requiring the energy in a Electromagnetic Calorimeter (EMCal) to match the measured
momentum of the tracks. Kaons are identified using the timing information from a Time Of Flight (TOF) detector and
EMCal which have very good $\pi$/K separation in the momentum range 0.3 $<p<$ 2.5 \gvcp and 0.3 $<p<$ 1.0 \gvcp,
respectively. Valid DC-PC1 tracks are confirmed by the matching of the associated hit information to the RICH and EMCal
in the case of electrons and to the TOF or EMCal in the case of kaons. 

The beam-beam counters (BBC) are used to determine the $z$-coordinate of the collision vertex ($z_{vtx}$) and in
combination with the zero-degree calorimeters (ZDC) provide the trigger and determine the event centrality.

In order to benefit from the high luminosity of RHIC and provide efficient detection of the rare electron and dielectron
events  PHENIX successfully implemented an electron trigger (ERT) which requires spatial matching between EMCal and RICH
and an energy above a certain threshold in the EMCal. 

\subsection{Electron pair analysis with PHENIX spectrometer}
\label{sec:2}
The electron pair analysis is performed with the sample of identified electrons using a statistical procedure in which all
particles in a given event are combined into pairs to generate unlike- and like-sign invariant mass spectra. By
construction the unlike-sign spectrum contains both the signal and an inherent combinatorial background of uncorrelated
pairs. The size and shape of the combinatorial background are determined using the event mixing technique in which the
particles from one event are combined with the particles from different events provided that all events belong to the
same centrality and vertex classes. 

The unlike-sign mixed event integral yield is normalized to the measured 2~$\times\sqrt{N^{++}N^{--}}$ yield. All yields
are calculated above \mee = 200 \mvce to exclude the correlated $e^{+}e^{+}$ and $e^{-}e^{-}$ pairs from double $\pi^0$
Dalitz decays and double conversions. PHENIX has refined the mixed-event technique to a very high precision of less than
$\pm$0.1\% which is confirmed by comparing the mixed event like-sign invariant mass spectra to the measured one. The
normalization procedure has been tested in four different approaches and found to be stable within
0.5\%~\cite{HQtoia}. Finally, the signal mass distribution is derived by subtracting the normalized mixed event spectrum
from the measured one.  

\section{$\phi$-meson production}
\label{sec:3}

The $\phi$-meson yield ($dN/dy$) and temperature ($T$) were derived from the invariant $m_T$-distribution: 
\begin{equation}
  \frac{1}{2 \pi m_{T}} \frac{d^{2}N}{dm_{T}dy}  = \frac{N^{\phi}_{raw}(m_{T}){\cdot}CF(m_T){\cdot}\epsilon_{trigger}(m_T)}
  {2\pi m_T{\cdot}N_{events}{\cdot}\epsilon_{emb}{\cdot}\epsilon_{rbr}{\cdot}BR{\cdot}\Delta m_{T}},
\end{equation}

where $N^{\phi}_{raw}(m_{T})$ is the raw $\phi$ yield,  $CF(m_T)$ is the correction factor to account for acceptance and
pair reconstruction efficiency, $\epsilon_{trigger}(m_T)$ is the electron trigger efficiency which is equal to one for
the analysis of minimum bias events, $N_{events}$ is the number of analyzed events, $\epsilon_{emb}$ is the pair embedding
efficiency which accounts for the reconstruction efficiency losses due to detector occupancy, $\epsilon_{rbr}$ is an
efficiency due to run-by-run variations of the detector performance. $BR$ is the branching ratio for
\phe or \kk and $\Delta m_{T}$ is the bin size. 

The raw $\phi$ yield,  $N^{\phi}_{raw}$, is derived by summing the content of the spectrum over a mass interval of $\pm
3\sigma_{tot}$ where $\sigma_{tot}$ is the total width calculated from the quadrature sum of the experimental mass
resolution and natural width of the $\phi$ meson.

The correction factor $CF(m_T)$ is determined using a Monte Carlo simulation. Single $\phi$ mesons were generated with
an exponential transverse momentum distribution.  
The $\phi$'s are decayed, propagated  through an emulator of the PHENIX detector and the resulting output is passed
through the whole analysis chain. For each $m_T$ bin the ratio of the generated yield to the reconstructed one gives the
correction factor $CF(m_T)$. 

Finally, the $dN/dy$ and $T$ are extracted from the corrected invariant $m_T$ spectra fitted with the following
exponential function, having $dN/dy$ and $T$ as parameters:
\begin{equation}
  \label{eq:eq_fit}
    \dfrac{1}{2\pi m_{T}}\dfrac{d^{2}N}{d m_{T} dy} = \dfrac{dN/dy}{2 \pi T (T+M_{\phi})} \exp{\dfrac{-(m_{T} -
        M_{\phi})}{T}},
\end{equation}
where $M_{\phi}$ is the PDG value of the $\phi$-meson mass.

\subsection{\dau collisions}
\label{sec:3dau}

\begin{table*}
  \caption{$dN/dy$ and $T$ for the \phe and \phk analysis in \dau collisions.}
  \label{tab:1}       
  \centering
  \begin{tabular}{lll}
    \hline\noalign{\smallskip}
    Decay channel & $dN/dy$ & $T$ ($MeV$)  \\
    \noalign{\smallskip}\hline\noalign{\smallskip}
    \phe & 0.056$\pm$0.015(stat)$\pm$0.028(syst) & 326$\pm$94(stat)$\pm$118(syst) \\
    \phk & 0.047$\pm$0.009(stat)$\pm$0.010(syst) & 414$\pm$13(stat)$\pm$~23(syst) \\
    \noalign{\smallskip}\hline
  \end{tabular}
\end{table*}

The measurements in \dau collisions establish the baseline information for possible nuclear matter effects and provide
an essential reference for the comparison to the measurements in \au collisions.

The presented results are based on about 31 million triggered \dau events with a threshold of 600 $MeV$
(see Sec.~\ref{sec:1a}) and about 62 million minimum-bias \dau events for the \phe and \phk analyses, respectively.

\begin{figure}
  \centering 
  \resizebox{0.34\textwidth}{!}{%
    \includegraphics{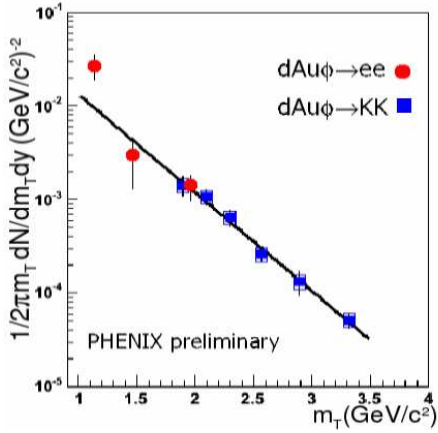}
  }
  \caption{(Color online) $m_T$ spectra of the $\phi$ mesons from the \phe and \phk analyses in \dau
    collisions. Statistical and systematic errors are shown by vertical bars and open rectangles, respectively.  The
    line represents the common fit to the exponential function, Eq.~\ref{eq:eq_fit}.}
  \label{fig:1} 
\end{figure}

The $\phi$-meson invariant $m_T$ spectra from both \ee and \kk analyses are shown in Fig.~\ref{fig:1} by the filled
circles and squares, respectively~\cite{seto}. The line represents the fit to all the data points with the exponential
function given by Eq.~\ref{eq:eq_fit}.  

$dN/dy$ and temperatures extracted from the \phe and \phk analyses in \dau collisions are listed in Table~\ref{tab:1}. 

\subsection{\au collisions}
\label{sec:3au}
In this section we present the $m_T$ spectra, invariant yield and temperature results for the different centralities
derived from the measurements of the $\phi$ meson through the \ee and \kk decay channels in \au collisions at \sqn = 200
$GeV$. The \phe analysis uses 903 million minimum-bias events. Kaons are combined in pairs using different combinations
of the TOF and EMCal used for the kaon identification (see Sec.~\ref{sec:1a}). The analysis performed with
409$\times$10$^6$ minimum-bias events for TOF-TOF and 170$\times$10$^6$ minimum-bias events for TOF-EMCal and
EMCal-EMCal detector combinations. 

\begin{figure}
  \centering 
  \resizebox{0.34\textwidth}{!}{%
    \includegraphics{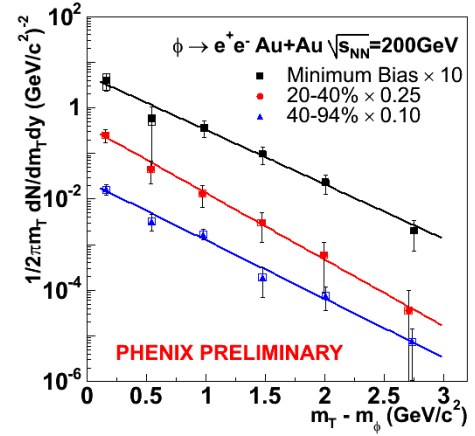}
  }
  \resizebox{0.34\textwidth}{!}{%
    \includegraphics{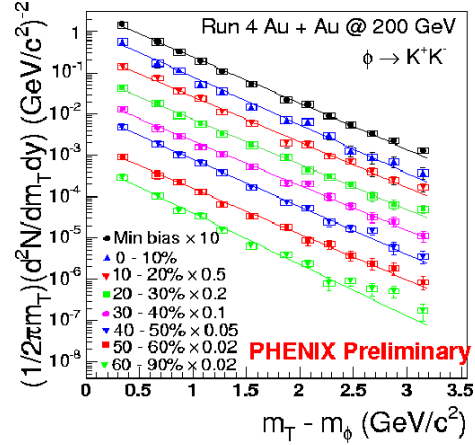}
  }
  \caption{(Color online) Invariant $m_T$ spectra of the \phe (top) and \phk (bottom) for minimum bias and several
    centrality bins in \au collisions. Statistical and systematic errors are shown by vertical bars and open rectangles,
    respectively. Each line represents the fit to the exponential function, Eq.~\ref{eq:eq_fit}.} 
  \label{fig:fig_invt} 
\end{figure}

The invariant $m_T$ spectra for minimum bias and several centrality bins, fitted with the exponential function
Eq.~\ref{eq:eq_fit}, are shown in the top and bottom panels of Fig.~\ref{fig:fig_invt} for \phe and \phk analyses,
respectively.

\begin{figure*}
  \centering 
  \resizebox{0.77\textwidth}{!}{%
    \includegraphics{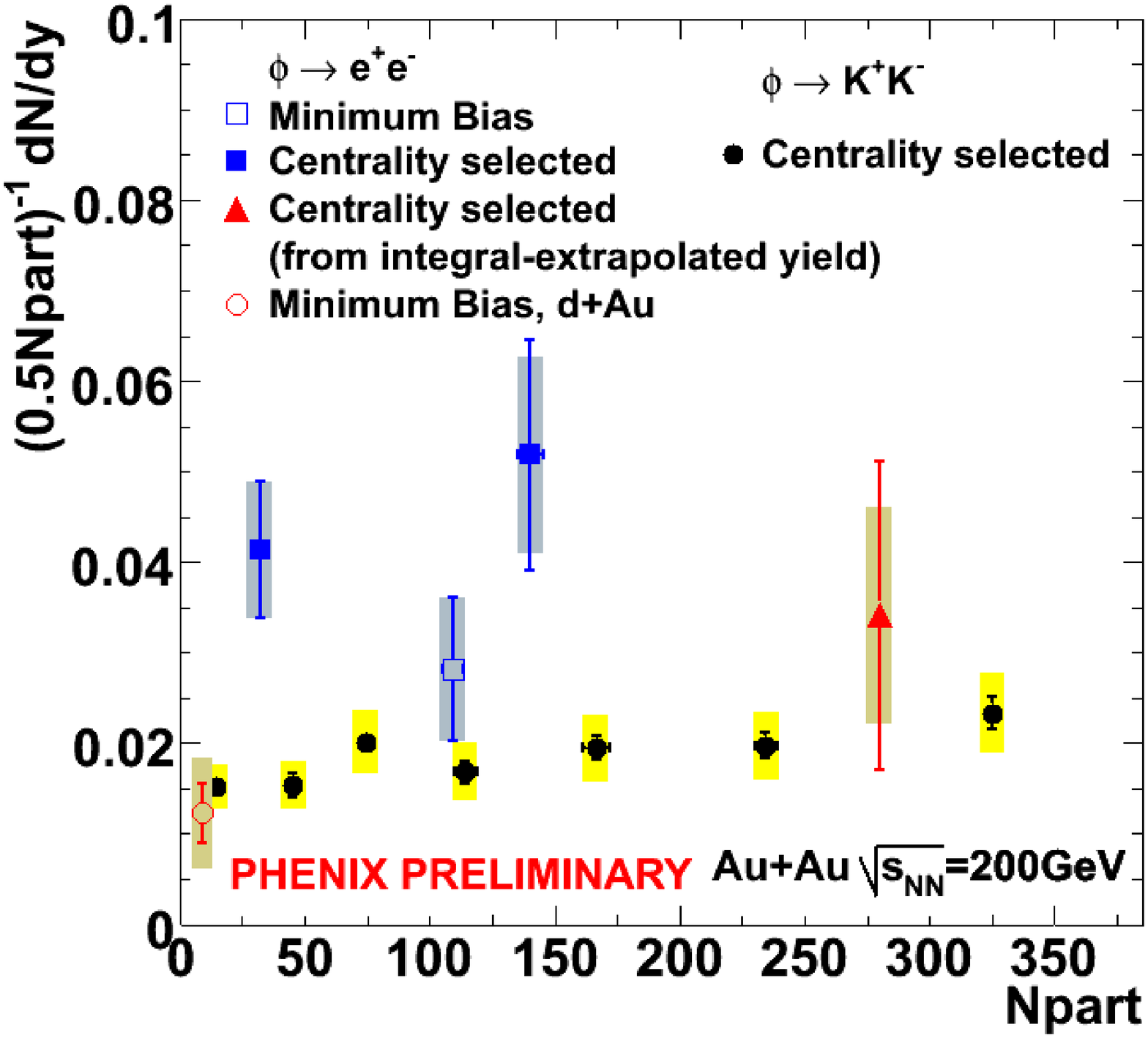}
    \hspace{3cm}
    \includegraphics{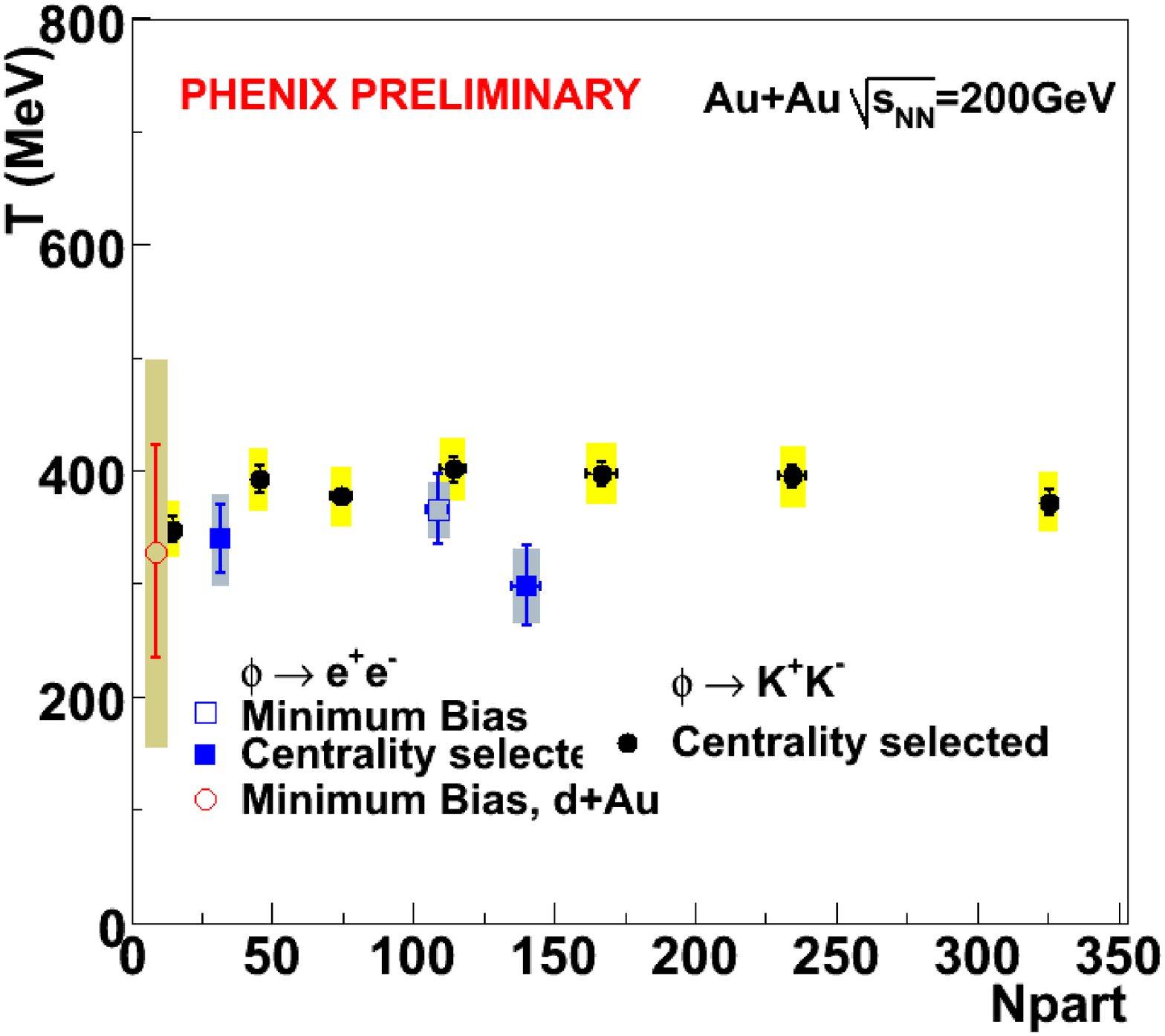}
    
  }
  \caption{(Color online) Multiplicity dependence of the $\phi$-meson yield normalized to number of participant pairs, (0.5$ \cdot
    N_{part}$)$^{-1}$ $dN/dy$ \ee and \kk decay channels. Open and filled squares represent minimum bias and centrality
    selected events, respectively. The triangle shows the $\phi$ yield derived from an independent analysis. The open
    circle represents the reference measurements of $dN/dy$ in \dau collisions. Statistical and systematic errors are
    shown by vertical bars and shaded bands, respectively.}
  \label{fig:fig_dndy_t} 
\end{figure*}

Fig.~\ref{fig:fig_dndy_t} shows the centrality dependence of the $\phi$-meson yield per participant pair 
extracted from the $m_T$ distributions in both analyses. The highest centrality bin shown by the triangle in the left
panel of Fig.~\ref{fig:fig_dndy_t} has limited statistics and is derived from an independent analysis. In this analysis
the invariant yield of the $\phi$ meson is obtained using the integral yield and the correction factor integrated over
$m_T$ assuming $T$ = 366 $MeV$.  

The yield per participant pair shows a $\sim$ 50\% increase in the \kk decay channel from peripheral to central
collisions. The present statistical and systematic uncertainties do not allow us to infer the centrality dependence of
the yield measured in the \ee decay channel (left panel of Fig.~\ref{fig:fig_dndy_t}). 

The comparison of the $\phi$-meson production measured via the \ee and \kk decay channels shown in
Fig.~\ref{fig:fig_dndy_t} may indicate a possible increase of the yield in the dielectron channel compared to the kaon
one. However, the statistical and systematic errors in the dielectron channel are too large for a definite
statement and within the error bars the yield in the two decay channels are consistent. The temperatures measured
in \au collisions through the \ee and \kk decay channels (see right panel of Fig.~\ref{fig:fig_dndy_t}) are centrality
independent and agree within the statistical and systematic uncertainties.

The open circles in Figs.~\ref{fig:fig_dndy_t} represent the reference measurements of the yield and $T$ in \dau
collisions and both are found to be in a good agreement with $dN/dy$ and $T$ measured in the peripheral \au collisions
via the \kk decay channel. The temperatures measured via the \ee decay channel in \dau and \au collisions are consistent
within the error bars. The yield per participant pairs measured in \dau collisions is about two times smaller than the one
measured in peripheral \au collisions in the \ee channel but the large statistical and systematic uncertainties do not allow for a
conclusive statement.  

\section{Low-mass dielectron continuum}
\label{sec:4}
The results presented in this section are based on the analysis of 800 million minimum-bias events collected in \au
collisions at \sqn = 200 $GeV$. 

\begin{figure}
  \centering 
  \resizebox{0.40\textwidth}{!}{%
    \includegraphics{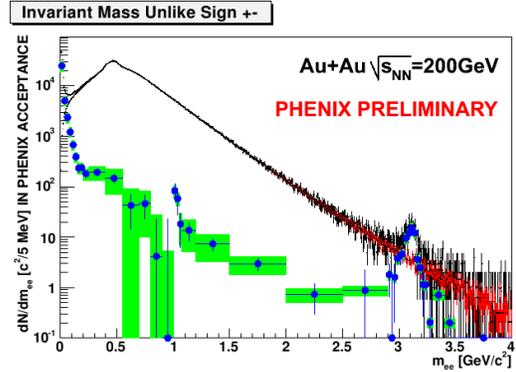}
  }
  \caption{(Color online) The foreground (black), background (red) and subtracted \ee invariant mass
    spectra. Statistical and systematic errors are shown by vertical bars and shaded bands, respectively.}
  
  \label{fig:fig_alb_log} 
\end{figure}

\begin{figure}
  \centering 
  \resizebox{0.39\textwidth}{!}{%
    \includegraphics{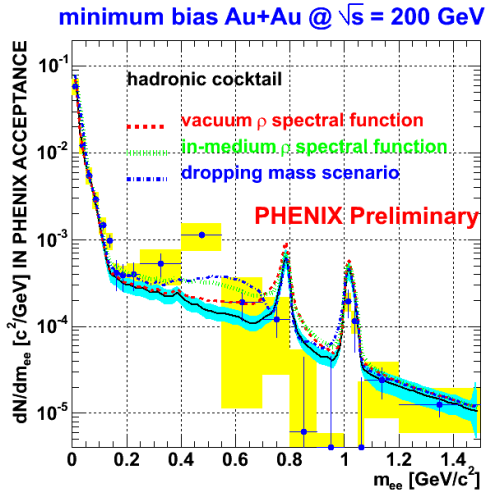}
  }
  \caption{(Color online) The measured \ee invariant mass spectrum compared to the hadronic cocktail (solid line) and
    calculations assuming the vacuum $\rho$ spectral function (dashed), in-medium broadening of the $\rho$ width
    (dotted) or $\rho$ dropping mass (dashed-dotted).} 
  \label{fig:fig_alb_th} 
\end{figure}

Fig.~\ref{fig:fig_alb_log} shows the foreground, background, and subtracted invariant mass spectra of the dielectron
pairs. The background distribution was obtained with an event mixing procedure with remarkable precision
(see Sec.~\ref{sec:2}). One can see that the signal to background ratio, $S/B$, at masses $m_{e^+e^-} \approx$ 400 \mvce
is of the order of $\sim$ 1/150. With such a small $S/B$ ratio the systematic errors, shown in Fig.~\ref{fig:fig_alb_log}
by shaded bands, are dominated by the normalization of the background spectrum.  

Fig.~\ref{fig:fig_alb_th} shows the comparison of the measured \ee invariant mass spectrum to the expected cocktail of
hadron decays and to the theoretical predictions from the $\pi\bar{\pi}$ annihilation channel without (dashed line) and
with (dotted and dashed-dotted lines) modification of the $\rho$-meson spectral function~\cite{rap02,rap01}. The results
shown in Fig.~\ref{fig:fig_alb_th} may indicate an enhancement of the dielectron yield in the mass range
between 0.2 to 0.6 \gvce over the expected hadronic cocktail and even the calculations involving in-medium modification
of the $\rho$ meson, although the present uncertainties do not allow us to draw a strong conclusion.

\begin{figure}
  \centering 
  \resizebox{0.37\textwidth}{!}{%
    \includegraphics{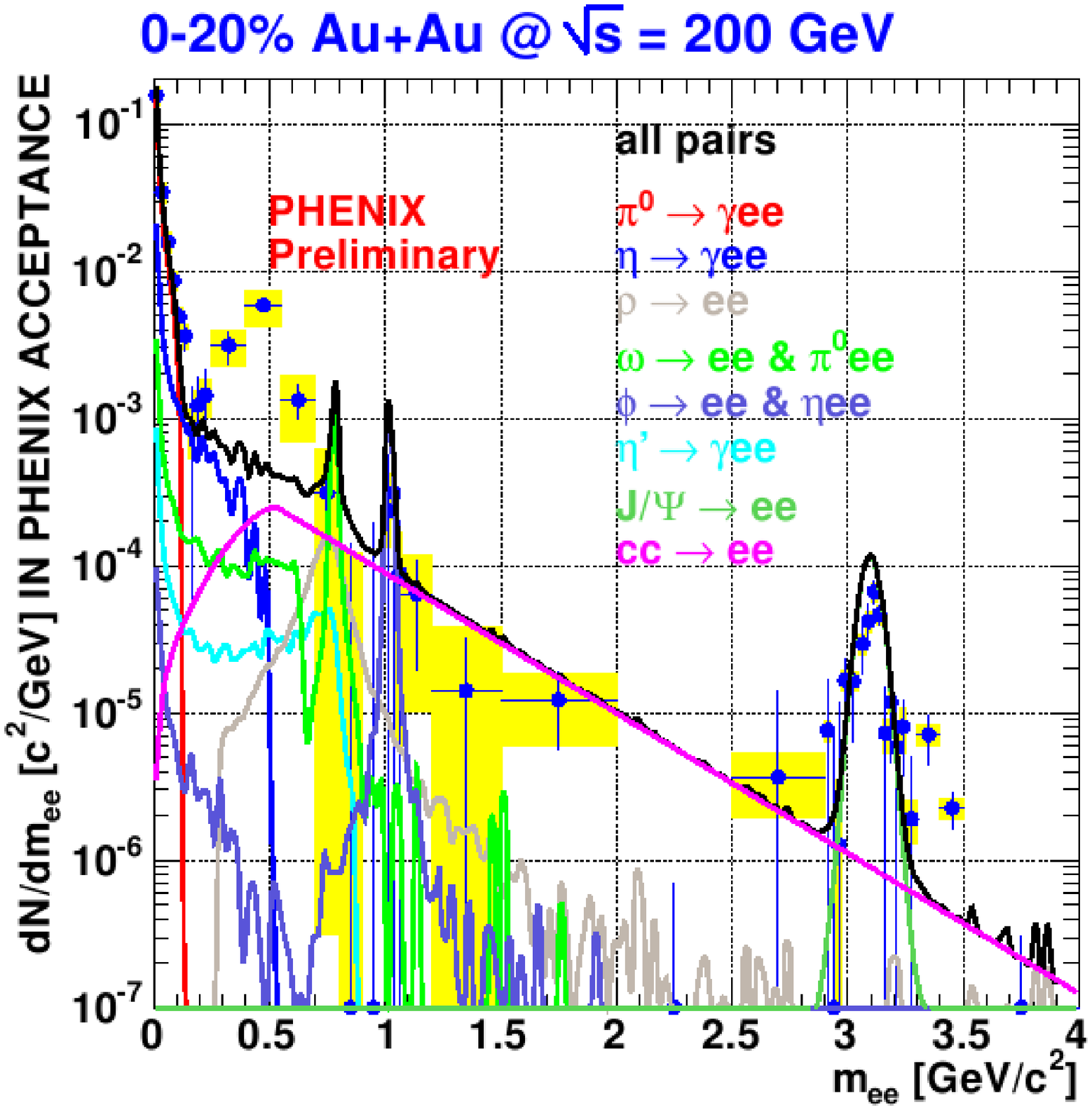}
  }
  \resizebox{0.37\textwidth}{!}{%
    \includegraphics{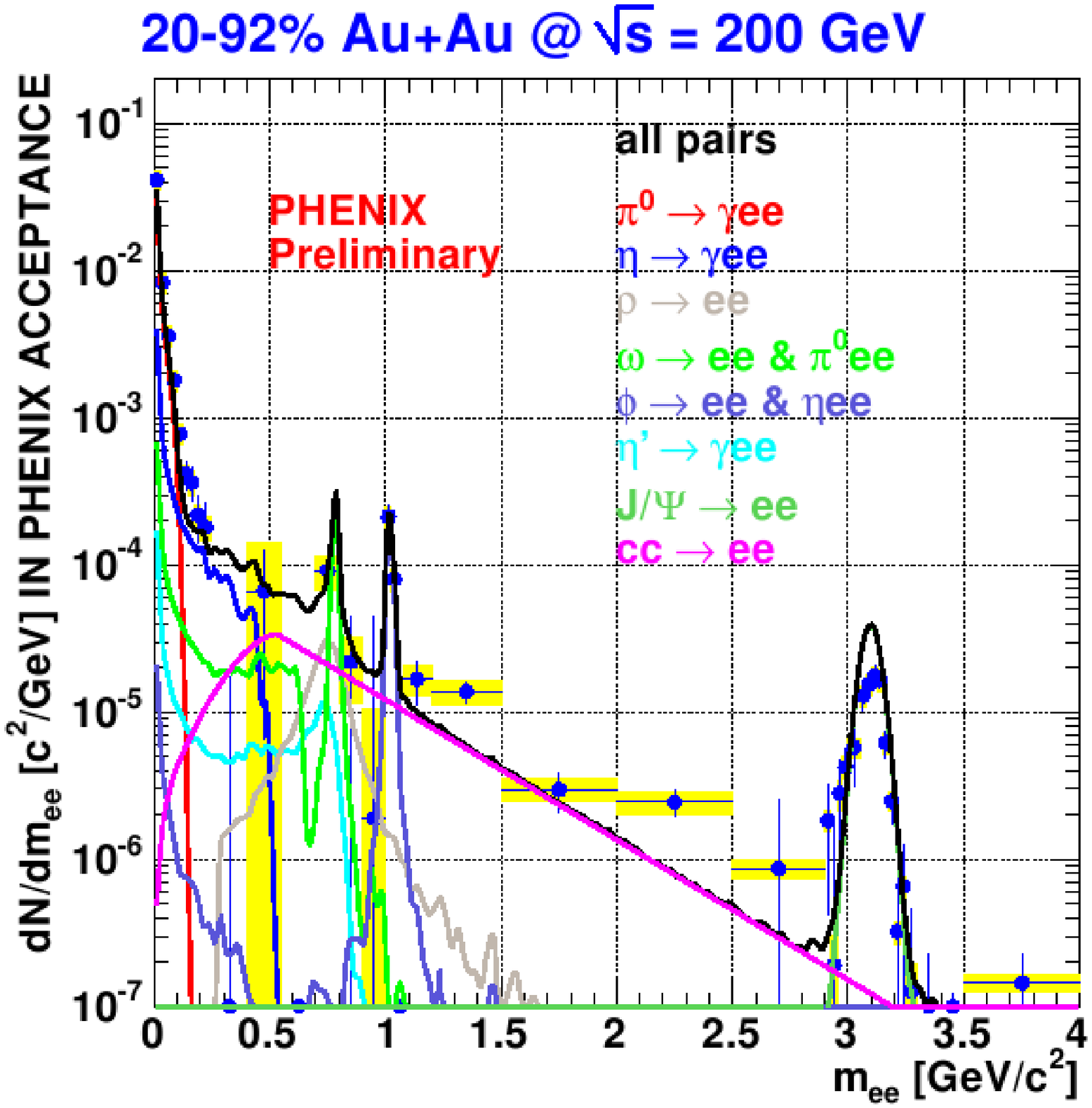}
  }
  \caption{(Color online) The comparison of the data to a hadronic cocktail and charm contribution for centrality
    classes 0-20\% (right) and 20-92\% (left). Statistical and systematic errors are shown by vertical bars and shaded
    bands, respectively.} 
  \label{fig:fig_alb_cock} 
\end{figure}

Fig.~\ref{fig:fig_alb_cock} shows the measured \ee spectra for the central (0-20\%) and semi-peripheral (20-92\%)
centrality classes compared to the hadronic cocktail. The possible excess of \ee pairs within mass region 0.2-0.6 \gvce
is seen in the central collisions whereas no excess is seen in the semi-peripheral ones.
\begin{figure*}
  \centering 
  \resizebox{0.89\textwidth}{!}{%
    \includegraphics{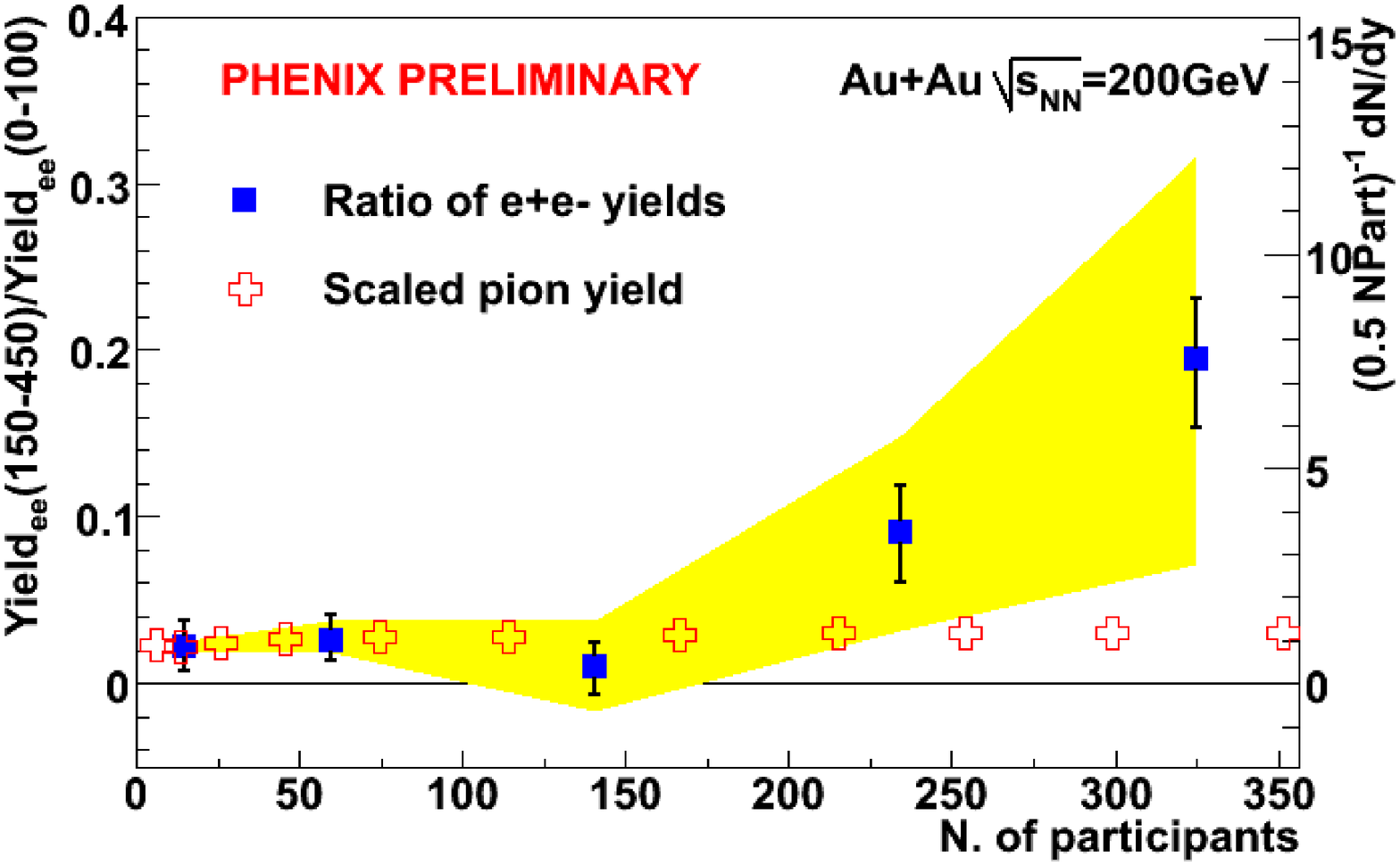}
    \hspace{2cm}
    \includegraphics{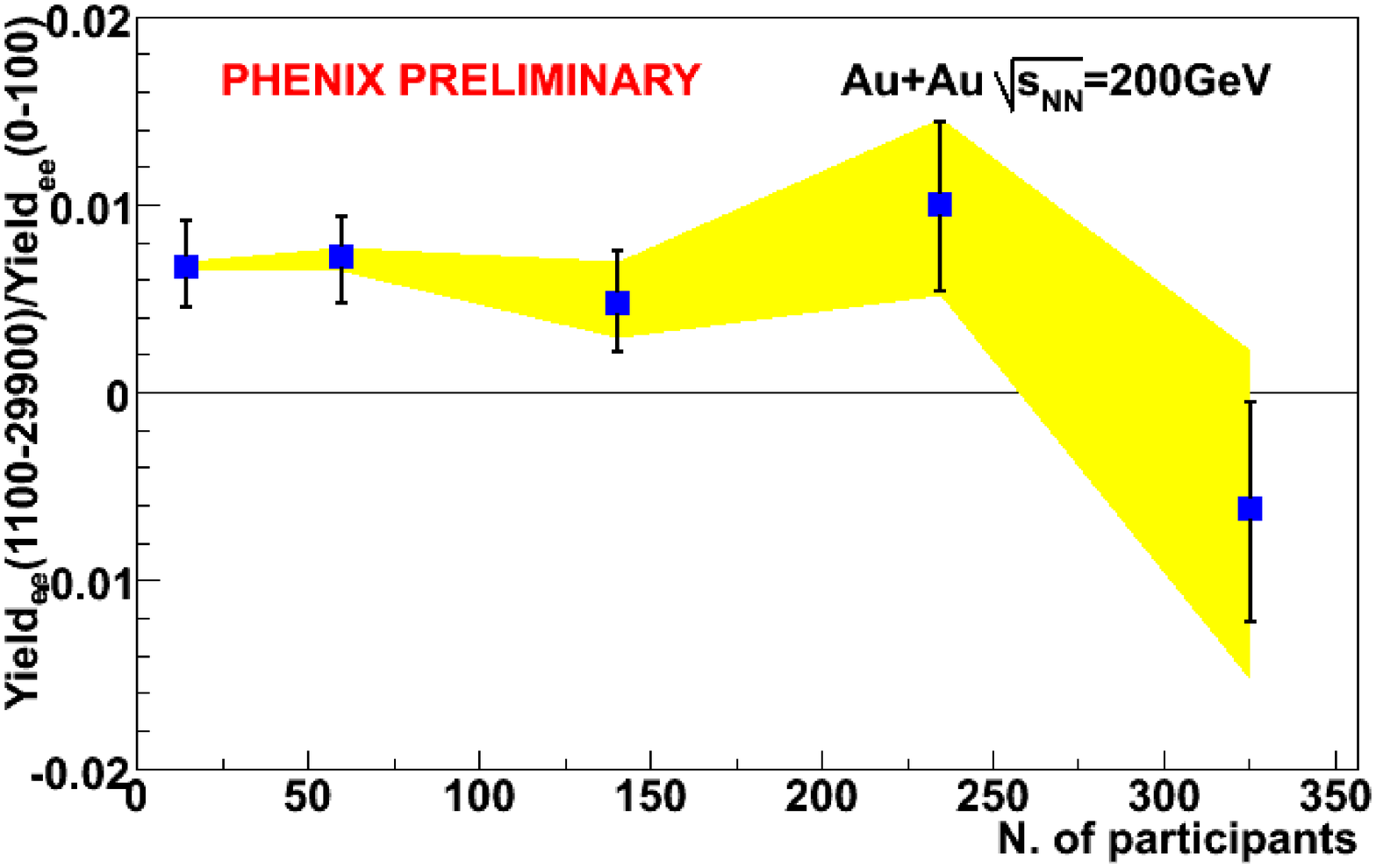}
  }
  \caption{Ratio of the dielectron yield measured in the mass region 150-450 \mvce (left) and 1.1-2.9 \gvce (right) with
    respect to the $\pi^0$ yield (\mee $<$~100 \mvce). Crosses represent the pion yield per participant pair as a
    function of the number of participants~\cite{pi0}. Statistical and systematic errors are shown by vertical bars and
    shaded bands, respectively.}
  \label{fig:fig_alb_ratio} 
\end{figure*}

The in-medium modification of the dielectron continuum can be studied also by looking at the ratio of the \ee yield in
different mass regions with respect to the $\pi^0$ Dalitz region (\mee $<$~100 \mvce) which is independent of the
number of participants. The left panel of Fig.~\ref{fig:fig_alb_ratio} shows the ratio of the yield in the mass region
between 150-450 \mvce overlaid with the pion yield per participant pair shown by the crosses~\cite{pi0}. The pion yield 
scales with $N_{part}$ while the ratio could have an indication of nonlinearity. The ratio for the mass region  1.2-2.9
\gvce (right panel) is within the uncertainties independent of $N_{part}$ although it is expected to increase following
the scaling with the number of participants of the charm production~\cite{charm} which is the main source of the
dielectrons in this mass region. The statistical significance of these results, however, is limited due to the low $S/B$
ratio and does not allow us to draw a conclusion.

\section{The Hadron Blind Detector}
\label{sec:5}

The capability of the PHENIX detector to measure low-mass dielectron pairs will be greatly improved with an
upgrade that will add a Hadron Blind Detector (HBD)~\cite{hbd1}. 

The detector is a conceptually new \v{C}erenkov detector operated with pure $CF_4$ in a proximity focus
configuration. It is coupled directly to a triple Gas Electron Multiplier ($GEM$)~\cite{sauli} detector with a $CsI$
photocathode layer evaporated on the top face of the first $GEM$ foil. The detector has a pad readout scheme. 

\begin{figure}
  \centering 
  \resizebox{0.39\textwidth}{!}{%
    \includegraphics{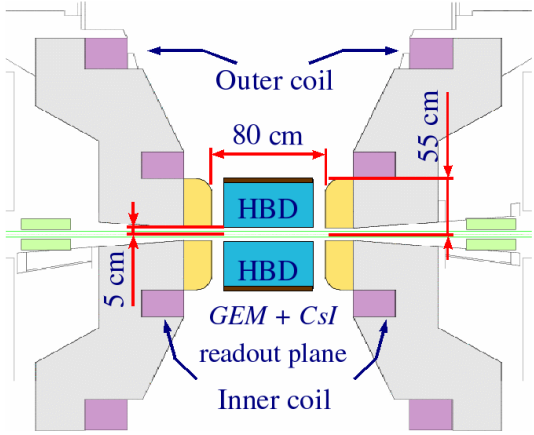}
  }
  \caption{Layout of the inner part of the PHENIX detector.}
  \label{fig:fig_hbd} 
\end{figure}

The HBD is located in the field free region extending up to $r \leq$ 60 cm of the inner part of the PHENIX detector
(Fig.~\ref{fig:fig_hbd}), which is realized by running the inner coil, recently installed in PHENIX for this purpose,
with opposite current to compensate for the field from the outer coil. \v{C}erenkov photons from an electron passing through
the radiator are directly collected on the $CsI$ photocathode. In this configuration
photons form  a circular blob, not a ring as in a RICH detector. The photoelectrons emitted from the photocathode are
amplified by the triple $GEM$ detector. Finally, the electron avalanche is read out in a pad plane located at the
bottom of the detector element.

The HBD exploits the fact that the opening angle of electron pairs from $\gamma$ conversions and $\pi^0$ Dalitz
decays is small compared to that of pairs from the light vector mesons. In the field-free region this angle is preserved
and by applying an opening angle cut one can reject more than 90$\%$ of the conversions and $\pi^{0}$ Dalitz decays,
while keeping most of the signal.

The rejection ability of the HBD is confirmed by realistic Monte-Carlo simulations which demonstrated that with the
HBD the combinatorial background originating from conversions and $\pi^0$ Dalitz decays is reduced by approximately two
orders of magnitude. 

A comprehensive R\&D program demonstrated the feasibility of the proposed concept\cite{hbd}.
The HBD construction and installation are nearing completion with first operation foreseen at the next RHIC
run~\cite{Tserruya:2006hx}. 

\section{Acknowledgments}

The author acknowledges support by the Israel Science Foundation, the MINERVA Foundation and the Nella and Leon Benoziyo
Center of High Energy Physics Research.

%
%

\end{document}